\documentclass[10pt]{article}
\usepackage{color}
\usepackage{geometry}
\usepackage{amsmath,amssymb,subfigure,comment}
\usepackage[algo2e,ruled,lined,titlenumbered]{algorithm2e}
\usepackage{graphicx}    

\newcommand{ \V }[1]{ \underline{#1} }
\newcommand{ \M }[1]{ \underline{\underline{#1}} }

\date{September 2009}

\title{Improved multiscale computational strategies for delamination}
\author{O. Allix, P. Gosselet and P. Kerfriden
\\
LMT-Cachan (ENS-Cachan/CNRS/UPMC/Pres UniverSud Paris)\\
61 avenue du Pr\'esident Wilson, 94235 Cachan, France \\
\texttt{\{allix,kerfriden,gosselet\}@lmt.ens-cachan.fr}}
\begin{document}

\maketitle

\section{Introduction}

The virtual testing of delamination is a goal shared by many
practitioners, especially in the aeronautical field. In order to
reach such an objective, two research topics which have undergone
drastic changes over the last twenty years must be linked: the
relevant modeling of composites and the efficient analysis of
structures.

Indeed, there have been many advances toward a better understanding
of the mechanics of laminated composites and the mechanisms of
damage. The validity of two types of models, microscale models and
mesoscale models, has been proven. Microscale models are closely
connected to the physics of the material and, thus, provide a
reliable framework for simulation. They take into account many
damage processes, such as diffuse intralaminar degradations
percolating into transverse cracking, diffuse interface degradations
leading to distributed delamination, chemically- or
thermally-induced degradations, or fiber breakage.
\cite{ladeveze05,degeorges07}. On the microscale, simulations can
combine continuous (damage) and discrete (fracture) degradation
models \cite{lubineau09}. Unfortunately, the analysis of models
defined on the microscale requires such a refined discretization
that only small test specimens can be simulated. Industrial-size
structural calculations are beyond the reach of even recent
computers. Mesomodels \cite{allix92,ladeveze02c,deborst06} are
defined on a scale which makes both the introduction of
physics-based components and the simulation of small industrial
structures possible. Very often these models rely on the definition
of two mesoconstituents, the ply (a three-dimensional entity) and
the interface (a two-dimensional entity), which are modeled using
continuum (damage) mechanics and behavior derived from the
homogenization of micromodels \cite{lubineau07,lubineau08}.
Nevertheless, in order to achieve reliable simulations, refined
discretizations are still required for the correct representation of
the stress gradients induced by edge effects, which are responsible
for the initiation of many degradations. Therefore, the resulting
problems remain very large (in terms of the number of degrees of
freedom) and highly nonlinear, which creates potential
instabilities.

In a first approach to the reliable simulation of delamination in
composite structures, we chose to neglect the effect of damage
within plies and concentrate on the degradations at the interfaces.
Thus, we adopted the mesomodel presented in \cite{allix92}, in which
the debonding phenomenon is localized at the interfaces and handled
through cohesive behavior. A similar approach with a different
interface behavior (degradations based on plasticity) was applied in
\cite{schellekens.1994.1}.

In order to handle the large nonlinear systems associated with this
modeling approach, one can consider using one of the several
multiscale \cite{mandel93,fish97,feyel00,ladeveze03b} and enrichment
\cite{oden99,hughes98,ghosh01,melenk96} techniques developed
recently. We based our strategy on the mixed domain decomposition
method described in \cite{ladeveze03b}, which places special
emphasis on the interfaces between substructures. Consequently, the
reference problem resulting from the mesomodel chosen is
substructured by nature, and the cohesive interfaces of the model
are handled within the interfaces of the domain decomposition
method. This idea is developed in
Section~\ref{sec:reference_problem}. Furthermore, the resolution of
the substructured problem by a LATIN iterative solver has very
advantageous numerical properties: the nonlinearities are dealt with
through local problems, and very little matrix reassembling is
required. The incremental micro-macro LATIN algorithm as a
resolution strategy for delamination problems is presented in
Section~\ref{sec:multiscale_resolution}. As shown in
Section~\ref{sec:delam_examp}, the direct application of this method
leads to a number of numerical difficulties. A first issue occurs
when setting the parameters of the method: in Section~\ref{sec:ddr},
we present the indispensable tuning of the search directions
according to the interface's status. Subsequently, the main
remaining difficulties concern the treatment of the macroscale of
the problem. In this paper, the emphasis is on the adaptation of our
strategy in order to deal with large macroproblems. (Important
remarks on how to make the macroproblem more relevant can be found
in \cite{kerfriden09}.) In Section~\ref{sec:third_scale}, we present
the parallelization of the resolution, which was inspired by
\cite{mandel93}. In order to do that, we introduce a third level of
discretization after the first and most refined level (the finite
element) and the second level (the substructure): interconnected
substructures are combined into ``super-substructures'' (which fill
up the memory capacity of processors) connected to one another
through ``super-interfaces'' using Message Passing Interface (MPI).
The method is validated in Section~\ref{sec:complex_test} using a
complex test case. The handling of the instabilities is not
described in this paper. The interested reader may refer to
\cite{kerfriden09} where the adaptation of an arc-length algorithm
with local control is presented.

\section{Application of the two-scale domain decomposition strategy to delamination analysis}
\label{sec:reference_problem}
\subsection{The substructured delamination problem}


Let us consider a laminated structure $\mathbf{E}$ defined in a
domain $\Omega$ bounded by $\partial \Omega$ and consisting of $N_P$
adjacent plies $P$, each defined in a domain $\Omega_P$, with
$\Omega = \bigcup_P \Omega_P$. Adjacent plies $P$ and $P'$ are
joined by cohesive interfaces $I_{PP'}$. An external traction field
$\V{F}_d$ is prescribed over a part $\partial \Omega_f$ of $\Omega$,
and a displacement field $\V{U}_d$ is prescribed over the complementary part 
$\partial \Omega_u$ of $\Omega$. Let $\V{n}_P$ denote the outer
normal to the boundary $\partial \Omega_P$ of Ply P, $\V{f}_d$ the
volume force, $\M{\sigma}$ the Cauchy stress tensor and
$\M{\epsilon}$ the symmetric part of the displacement gradient. The
simulation is performed using a classical incremental scheme,
assuming small perturbations and quasi-static isothermal evolution
over time.


\begin{figure}[htb]
       \centering
       \includegraphics[width=0.9 \linewidth]{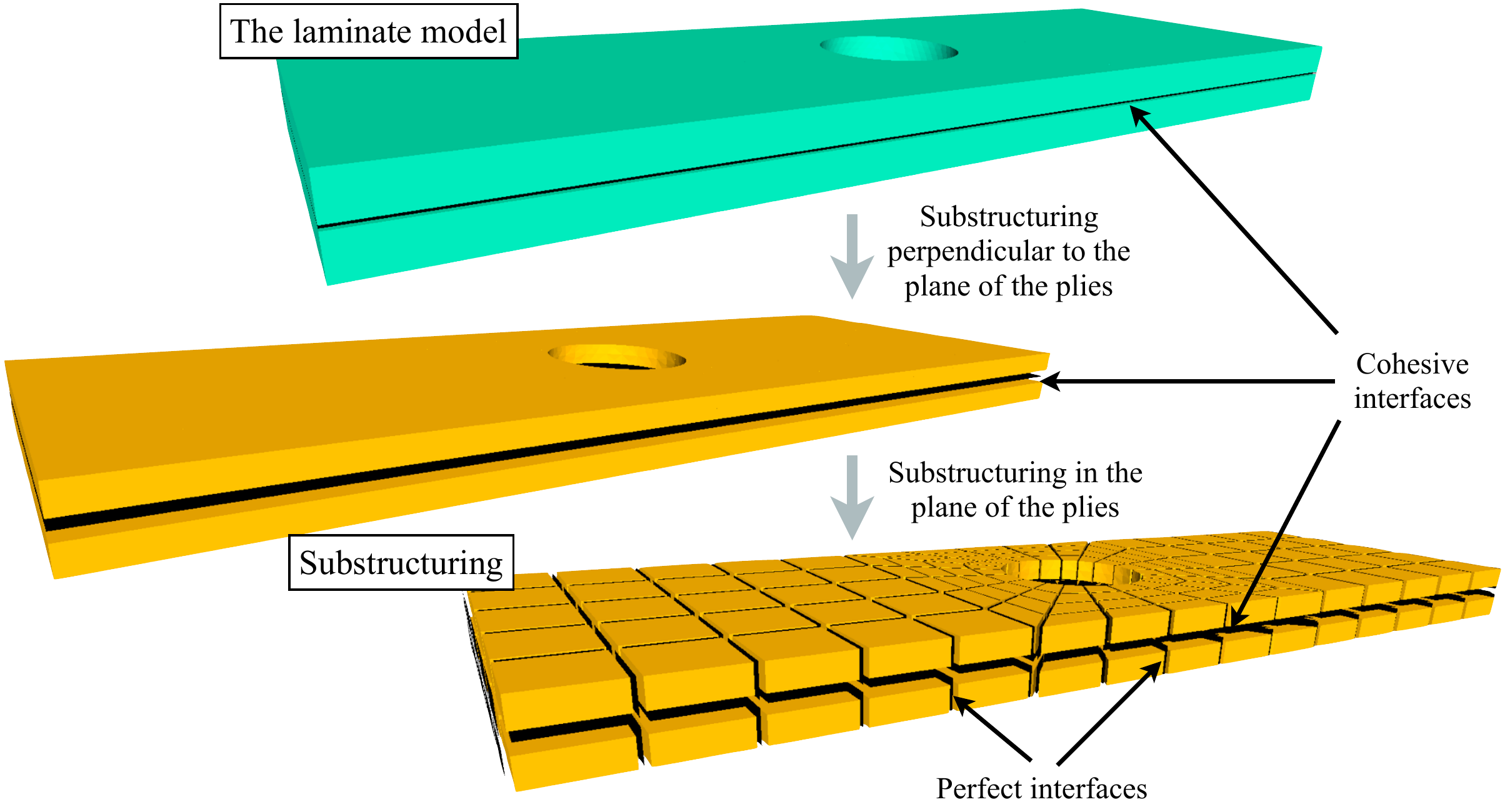}
       \caption{Decomposition of the laminated composite structure into substructures}
       \label{fig:decomp_sst_interfaces}
\end{figure}

\begin{figure}[htb]
       \centering
       \includegraphics[width=0.7 \linewidth]{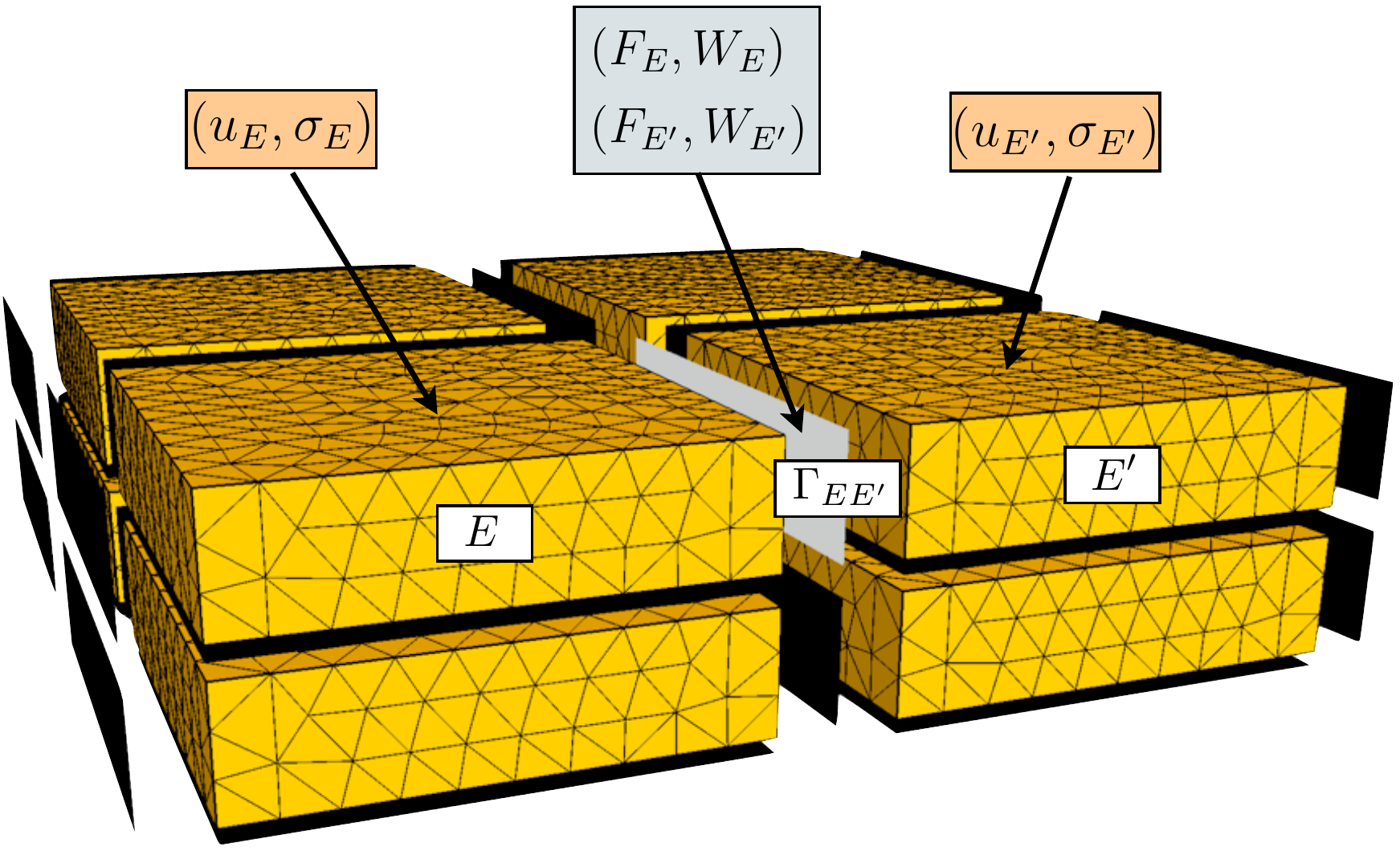}
       \caption{Decomposition of the laminated composite structure into substructures:
       mixed unknown fields}
       \label{fig:champs_interface}
\end{figure}

The laminated structure $\mathbf{E}$ is decomposed into
substructures and interfaces as shown in
Fig.~\ref{fig:decomp_sst_interfaces}. Each of these mechanical
entities has its own kinematic and static unknown fields as well as
its own constitutive law. The substructuring pattern is defined in
such a way that the domain decomposition interfaces coincide with
the material's cohesive interfaces, so that each substructure
belongs to a unique ply $P$ and has a constant linear constitutive
law. A substructure $E$ defined in Domain $\Omega_E$ is connected to
an adjacent substructure $E'$ through an interface
$\Gamma_{EE'}=\partial \Omega_E \cap\partial \Omega_{E'}$
(Fig.~\ref{fig:champs_interface}). The surface entity $\Gamma_{EE'}$
applies force distributions $\V{F}_E$, $\V{F}_{E'}$ and displacement
distributions $\V{W}_E$, $\V{W}_{E'}$ to $E$ and $E'$. Let $\Gamma_E
= \bigcup_{E' \in \mathbf{E}} \Gamma_{EE'} $. Over a substructure
$E$ such that $\Gamma_E \cap\partial \Omega\neq \emptyset$, the
boundary condition $(\V{U}_d,\V{F}_d)$ is applied through a boundary
interface $\Gamma_{E_d}$.

At each step of the incremental time resolution algorithm, the
substructured quasi-static problem consists in finding $s= (s_E)_{E
\in \mathbf{E}} $ (where $s_E = (\V{u}_E , \V{W}_E , \M{\sigma}_E ,
\V{F}_E )$) which is a solution of the following equations:
\begin{itemize}
\item kinematic admissibility of Substructure $E$:
\begin{equation}
\textrm{at each point of } \Gamma_E, \quad {\V{u}_E} = {\V{W}_{E}}
\end{equation}
\item static admissibility of Substructure $E$:
\begin{equation}
\label{equation:equilibre_ss}
 \begin{array}{l}
\displaystyle \forall ({\V{u}_E}^\star,{\V{W}_E}^\star) \in \mathcal{U}_{E} \times \mathcal{W}_{E} \  / \  {{\V{u}_E}^\star}_{| \partial \Omega_E} = {{\V{W}_{E}}^\star},  \\
\begin{array}{ll}
\displaystyle \int_{\Omega_E} Tr \left( \M{\sigma}_E \, \M{\epsilon} ({\V{u}_E}^\star) \right) \ d & \Omega = \displaystyle \int_{\Omega_E} \V{f}_d . {\V{u}_E}^\star \, d \Omega \\
&+\displaystyle  \int_{\partial{\Gamma_E}} \V{F}_E . {\V{W}_E}^\star \ d \Gamma
\end{array}
\end{array}
\end{equation}
\item linear orthotropic constitutive law of Substructure $E$:
\begin{equation}
\textrm{at each point of } \Omega_E,\quad \M{\sigma}_E = \mathbf{K} \, \M{\epsilon}(\V{u}_E)
\end{equation}

\item behavior of the interfaces $\Gamma_{EE'} \in \Gamma_E$:
\begin{equation}
\begin{array}{l}
\displaystyle \textrm{at each point of } \Gamma_{EE'} \in \Gamma_E, \\
\displaystyle R_{EE'}( \V{W}_{E}  , \V{W}_{E'} , \V{F}_{E} , \V{F}_{E'} ) = 0
\end{array}
\end{equation}
\item behavior of the interfaces at the boundary $\partial \Omega \cap \Gamma_E$:
\begin{equation}
\begin{array}{c}
\textrm{at each point of } \Gamma_{{E}_d}, \quad R_{Ed}( \V{W}_{E}, \V{F}_{E}) = 0\\
(\V{W}_{E}=\V{u}_d \text{ on } \partial\Omega_u \text{ and } \V{F}_{E}=\V{F}_d \text{ on }\partial\Omega_f)
\end{array}
\end{equation}
\end{itemize}

\begin{figure}[htb]
       \centering
       \includegraphics[width=0.8 \linewidth]{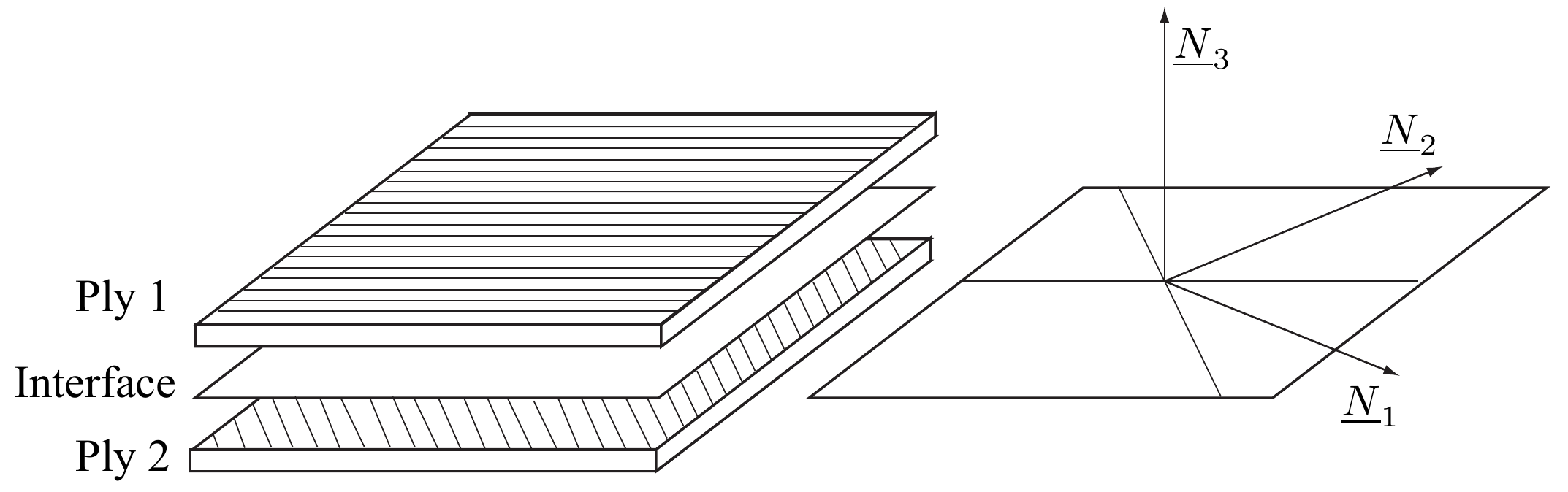}
       \caption{The components of the mesomodel}
       \label{fig:interface_composite}
\end{figure}

We make the formal relation $R_{EE'}=0$ explicit in the two cases we
will be considering:
\begin{itemize}
\item perfect interface:
\begin{equation}
\left\{ \begin{array}{l}
\V{F}_{E} + \V{F}_{E'} = 0 \\
\V{W}_{E} - \V{W}_{E'} = 0
\end{array} \right.
\end{equation}
\item cohesive interface:
\begin{equation}
\left\{ \begin{array}{l}
\displaystyle \V{F}_{E} = \M{K}_P(\V{[W]}_{EE'}(\tau<t)). \V{[W]}_{EE'} \\
\displaystyle \V{F}_{E} + \V{F}_{E'} = 0 \\
\end{array} \right.
\end{equation}
where $\V{[W]}_{EE'} = \V{W}_{E'}-\V{W}_{E}$. The stiffness
operator $\M{K}_P$ can be expressed in the
$(\V{N}_1,\V{N}_2,\V{n}_P)$ basis as
(Fig.~\ref{fig:interface_composite}):
\begin{equation}
\left( \begin{array}{ccc}
\displaystyle (1-d_1){k_1}^0 & 0 & 0 \\
0 & \displaystyle (1-d_2){k_2}^0 & 0 \\
0 & 0 & \displaystyle \left(1-h_+\V{[W]}_{EE'}\ d_3 \right){k_3}^0
\end{array}
\right)
\end{equation}
where $h_+$ is the positive indicator function.
\end{itemize}

The cohesive constitutive law of an interface joining two adjacent
plies is described classically through continuum damage mechanics.
Local damage variables $({d_i})_{i \in [1 \ 3]}$, with values
ranging from $0$ (healthy interface) to $1$ (completely damaged
interface), are introduced into the interface model in order to
simulate its progressive softening. The parameters $(d_i)_{i \in [1
\ 3]}$ are related to the local energy release rates $(Y_i)_{i \in
[1 \ 3]}$ of the interface's degradation modes (traction along
Direction $\V{N}_3$ and shear along Directions $\V{N}_1$ and
$\V{N}_2$).
\begin{equation}
Y_i = - \frac{\partial e_d}{\partial d_i}
\qquad \textrm{where} \quad
\left\{ \begin{array}{ccl}
Y_1 & = & \displaystyle \frac{1}{2} k_1 \left(\V{[W]}_{EE'}.\V{N}_1\right)^2 \\
Y_2 & = & \displaystyle \frac{1}{2} k_2 \left(\V{[W]}_{EE'}.\V{N}_2\right)^2 \\
Y_3 & = & \displaystyle \frac{1}{2} k_{3+} \left(\V{[W]}_{EE'}.\V{N}_{3}\right)_+^2
\end{array} \right. \end{equation}
The damage variables are assumed to be functions of a single
quantity: the maximum over time $Y_{|t}$ of a combination of the
energy release rates ${({Y_i}_{|\tau})}_{i \in [1 \ 3], \ \tau
\leqslant t}$:
\begin{equation}
Y_{|t} =  sup_{(\tau \leq t)} \left( {Y_3}_{|\tau}^\alpha + \gamma_1 {Y_1}_{|\tau}^\alpha + \gamma_2 {Y_2}_{|\tau}^\alpha \right)^{\frac{1}{\alpha}}
\end{equation}
The evolution laws express that:
\begin{equation}
d_1 = d_2 = d_3 = w(Y) \quad\textrm{where, in general,} \quad w(Y) = \frac{n}{n+1} \left( \frac{Y}{Y_c} \right)^n
\end{equation}
$n$ and $\alpha$ being scalar parameters of the model. When
Parameters $\gamma_1$ and $\gamma_2$ are set to identified physical
values such that $\gamma_1 \neq \gamma_2 \neq 1$, the energies
dissipated during the propagation of the crack are different for the
three modes. Details on the identification procedure for such a
model can be found in \cite{ALLIX.1998.10}.

After a cohesive interface has become fully damaged, it is converted
into a (frictionless) contact interface.

\subsection{Two-scale iterative resolution of the substructured problem}
\label{sec:multiscale_resolution}

\subsubsection{Introduction of the macroscopic scale}

In the end, the substructured problem defined in the previous
section will be solved using an iterative LATIN algorithm, which
will be described in the next section. In order to ensure the
scalability of the strategy, a coarse global problem, associated
with the equilibrium and continuity of what one calls the ``macro''
force and displacement fields at the interfaces, must be solved at
each iteration.

Over each interface $\Gamma_{EE'}$ such that $(E,E') \in
\mathbf{E}^2$, the interface fields are divided into a macro part
(superscript $M$) and a micro part (superscript $m$). The macro part
belongs to a small subspace (9 macro degrees of freedom per plane
interface for a 3D problem).
\begin{equation}
\begin{array}{l}
\displaystyle {\V{F}_E} = \V{F}_E^M + \V{F}_E^m \\
\displaystyle {\V{W}_E} = \V{W}_E^M + \V{W}_E^m
\end{array}
\end{equation}
The macro and micro data are uncoupled with respect to the
interface's virtual work:
\begin{equation}
\begin{array}{l}
\displaystyle \forall \displaystyle (\V{F}_E,\V{W}_E)\in\mathcal{F}_E\times\mathcal{W}_E, \quad
\\ \displaystyle \int_{\Gamma_{EE'}} \V{F}_E.\V{W}_E \ d \Gamma
=  \int_{\Gamma_{EE'}} \V{F}_E^M.\V{W}_E^M \ d \Gamma
  + \int_{\Gamma_{EE'}} \V{F}_E^m.\V{W}_E^m \ d \Gamma
\end{array}
\end{equation}
Each macrospace is defined by one's choice of its basis. Numerical
tests have shown that the use of a linear macro basis gives the
method good scalability properties. Indeed, the corresponding
macrospace includes the part of the interface fields with the
largest wavelength. Consequently, according to Saint-Venant's
principle, the micro complement resulting from the iterative
resolution of the local problems has only a local influence.

\subsubsection{The iterative algorithm}

Here, the iterative LATIN algorithm for the resolution of nonlinear
problems is applied to the resolution of the substructured reference
problem, the nonlinearities being localized in the (cohesive) interfaces.\\

The equations of the problem can be divided into two groups:
\begin{itemize}
    \item linear equations in substructure variables and interface macroscopic variables:
    \begin{itemize}
        \item static admissibility of the substructures
        \item kinematic admissibility of the substructures
        \item linear constitutive law of the substructures
        \item linear equilibrium of the macro interface forces
    \end{itemize}
    \item local equations in interface variables:
    \begin{itemize}
        \item behavior of the interfaces
    \end{itemize}
\end{itemize}

The solutions $\displaystyle s = (s_E)_{E \in \mathbf{E}} = (\V{u}_E
, \V{W}_E , \M{\sigma}_E , \V{F}_E )_{E \in \mathbf{E}}$ of the
first set of equations belong to Space $\mathbf{A_d}$ and the
solutions $\displaystyle \widehat{s} = (\widehat{s}_E)_{E \in
\mathbf{E}} = (\V{\widehat{u}}_E , \V{\widehat{W}}_E ,
\M{\widehat{\sigma}}_E , \V{\widehat{F}}_E )_{E \in \mathbf{E}}$ of
the second set of equations belong to $\boldsymbol{\Gamma}$. The
converged solution $s_{ref}$ is such that:
\begin{equation}
s_{ref} \in \mathbf{A_d} \bigcap \boldsymbol{\Gamma}
\end{equation}

\begin{figure}[htb]
       \centering
       \includegraphics[width=0.6 \linewidth]{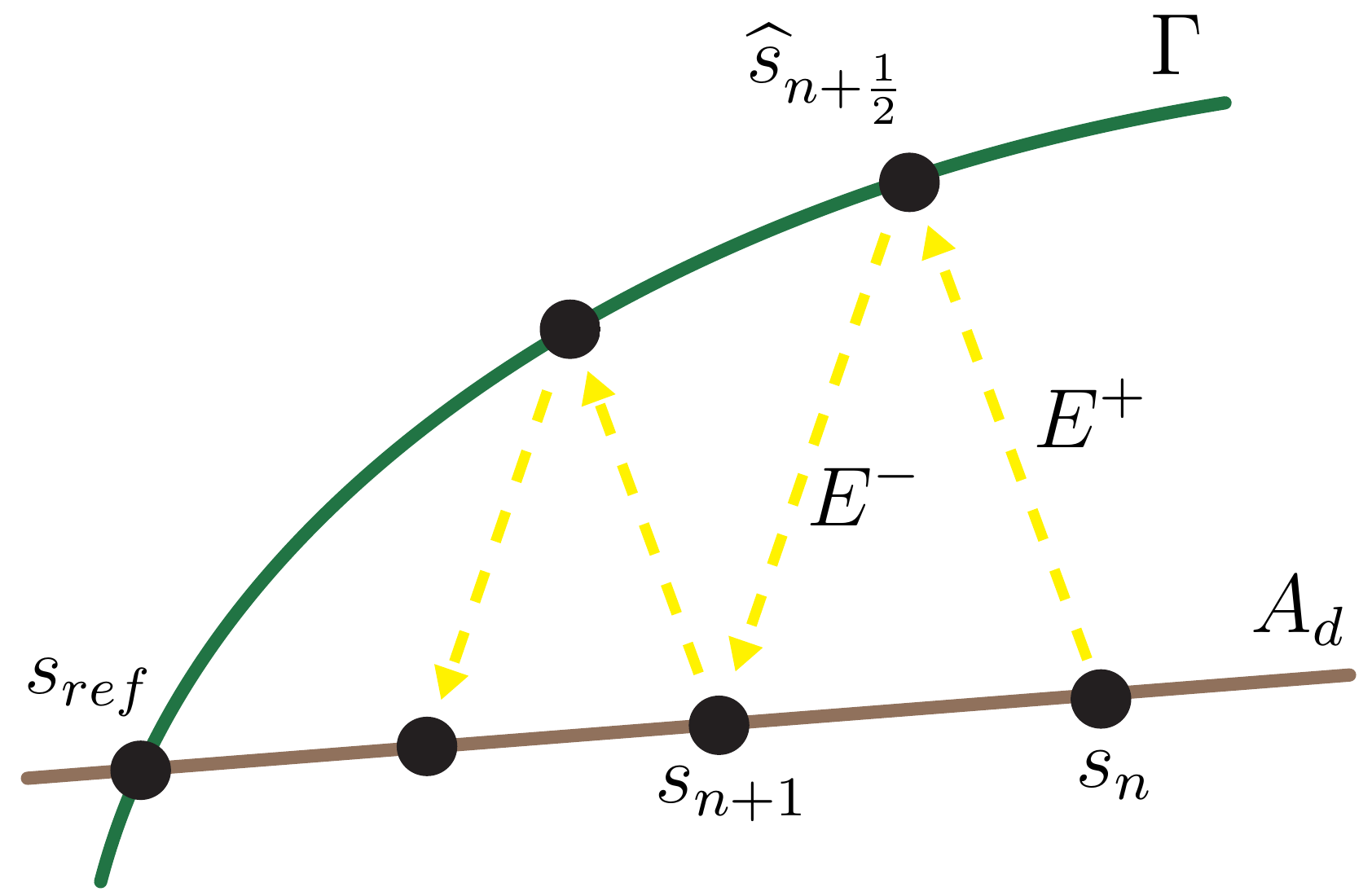}
       \caption{Illustration of the LATIN iterative algorithm}
       \label{fig:latin}
\end{figure}

The resolution process consists in seeking the solution $s_{ref}$
alternatively in these two spaces: first, a solution $s_n$ is found
in $\mathbf{A_d}$, then a solution $\widehat{s}_{n+\frac{1}{2}}$ is
found in $\boldsymbol{\Gamma}$. In order for the two problems to be
well-posed, one introduces two search directions, $\mathbf{E^+}$ and
$\mathbf{E^-}$, linking the solutions $s$ and $\widehat{s}$ through
the iterative process (see Fig. \ref{fig:latin}). Hence, an
iteration of the LATIN algorithm consists of two stages:
\begin{itemize}
\item a local stage:
\begin{equation}
\textit{Find} \ \widehat{s}_{n+\frac{1}{2}} \in \boldsymbol{\Gamma} \ \textit{such that}  \  \left( \widehat{s}_{n+\frac{1}{2}} -s_{n} \right) \in \mathbf{E^+}
\end{equation}
\item and a linear stage:
\begin{equation}\label{eq:ddrEm}
\textit{Find} \ s_{n+1} \in \mathbf{A_d} \ \textit{such that}  \  \left( s_{n+1} - \widehat{s}_{n+\frac{1}{2}} \right) \in \mathbf{E^-}
\end{equation}
\end{itemize}
In the following sections, the subscript $n$ will be omitted.

\paragraph{The local stage}

During the local stage, uncoupled problems are solved at each point
of the interfaces $(\Gamma_{EE'})_{|(E,E') \in \mathbf{E}^2}$ (as
well as $(\Gamma_{E_d})_{E \in \mathbf{E}}$ for the interfaces which
belong to the boundary $\partial \Omega$):
\begin{equation}
\label{eq:local_problem}
\begin{array}{l}
\textit{Find} \ (\V{\widehat{F}}_E,\V{\widehat{W}}_E,\V{\widehat{F}}_{E'},\V{\widehat{W}}_{E'})  \vspace{0.2cm}
\textit{ such that: } \\
\left\{ \begin{array}{l}
\displaystyle \mathcal{R}_{EE'}(\V{\widehat{F}}_E,\V{\widehat{W}}_E,\V{\widehat{F}}_{E'},\V{\widehat{W}}_{E'}) = 0  \\
\displaystyle  (\V{\widehat{F}}_E-\V{F}_E) - k^+ (\V{\widehat{W}}_E - \V{W}_E) = 0 \\
\displaystyle  (\V{\widehat{F}}_{E'}-\V{F}_{E'}) - k^+ (\V{\widehat{W}}_{E'} - \V{W}_{E'}) = 0
\end{array} \right.
\end{array}
\end{equation}
The last two equations of this system define the search direction
$E^+$ ($k^+$ and $k^-$ are scalar search direction which,
physically, are analogous to ``stiffnesses"). In the case of a
cohesive interface, Problem \eqref{eq:local_problem} is nonlinear
and its solution is obtained through a Newton-Raphson scheme.


\paragraph{The linear stage}

The linear stage consists in the resolution of a series of linear
systems within the substructures under the constraint of macroscopic
equilibrium of the interface forces.
\begin{equation}
\label{eq:macro_ad}
\text{at Interface }{\Gamma_{EE'}}_{|(E,E') \in \mathbf{E}^2}, \quad \V{F}_E^M + \V{F}_{E'}^M =  0
\end{equation}
In order to verify the macroscopic condition exactly and the search
direction $E^-$ defined in \eqref{eq:ddrEm} as well as possible, we
use a Lagrangian formulation, which leads to:
\begin{equation} \label{eq:ddr_loc}
\begin{array}{ll}
\displaystyle \forall  \V{W}_E^\star \in \mathcal{W}_E, & \displaystyle \quad \int_{\Gamma_E} \left( \V{F}_{E}-   \V{\widehat{F}}_{E} \right)  . {\V{W}_E}^\star \ d\Gamma
 \\
& \displaystyle + \int_{\Gamma_E}  \left( k^- \ (\V{W}_{E}-\V{\widehat{W}}_{E})   - k^- \V{\widetilde{W}}^M \right) . {\V{W}_E}^\star \ d\Gamma= 0
\end{array}
\end{equation}
which can be viewed as a modified search direction, the Lagrange
multiplier $\V{\widetilde{W}}^{M}$ becoming an additional unknown of
Interface $\Gamma_{EE'}$.

The problem which needs to be solved for each substructure $E$ is
obtained by substituting \eqref{eq:ddr_loc} into
\eqref{equation:equilibre_ss}:
\begin{equation}
\label{eq:pb_micro}
\begin{array}{l}
\displaystyle \forall (\V{u}_E^\star,\V{W}_E^\star) \in {\mathcal{U}_E} \times {\mathcal{W}_E},
\\
\displaystyle  \int_{\Omega_E} Tr (\M{\epsilon} (\V{u}_E) \,  K \M{\epsilon} ({\V{u}_E}^\star) ) \ d\Omega + \int_{\Gamma_E} k^- \, \V{W}_E . {\V{W}_E}^\star \ d \Gamma
 \\
\displaystyle =  \int_{\Omega_E} \V{f}_d . {\V{u}_E}^\star \, d \Omega + \int_{\Gamma_E} (\V{\widehat{F}}_E +k^- \V{\widehat{W}}_E +k^- {\V{\widetilde{W}}}^M) . {\V{W}_E}^\star \ d \Gamma
\end{array}
\end{equation}
The condensation of this equation onto the macro degrees of freedom
leads to a relation between $\V{F}_{E}^M$ and
$\V{\widetilde{W}}_{E}^M$ which can be introduced into the macro
equilibrium equation \eqref{eq:macro_ad}. Finally, one gets a small
linear system defined in the macro degrees of freedom. All the
subdomains contribute to that ``global'' system through homogenized
(condensed) flexibilities $\mathbb{L}_E^M$ calculated explicitly.
\begin{equation}
\label{eq:pb_macro}
\begin{array}{ll}
\displaystyle \forall & \displaystyle \V {\widetilde{W}}^{M \star} \in {\mathcal{W}^M}, \quad
\sum_E \int_{\Gamma_{E}} \mathbb{L}_E^M \ \V{\widetilde{W}}^M . \V {\widetilde{W}}^{M\star} \ d\Gamma
 \\ & \displaystyle =
 \sum_E \int_{\partial \Omega_f} \V{F}_d .\V{\widetilde{W}}^{M\star} \ d\Gamma
 -  \sum_E \int_{\Gamma_{E}}  \V{\widetilde{F}}_E .\V{\widetilde{W}}^{M\star} \ d\Gamma
\end{array}
\end{equation}
The macroscopic problem is discrete by nature and is expressed in
matrix form as $\displaystyle \mathbf{L}^M \ \widetilde{W}^M = F^M$,
where $\widetilde{W}^M$ is the vector of the components of the
Lagrange multiplier in the macro basis.

The right-hand side of Equation \eqref{eq:pb_macro} can be viewed as
a macroscopic static residual from the calculation of a single-scale
linear stage. In order to derive this term, Problem
\eqref{eq:pb_micro} must be solved independently within each
substructure. The resolution of the macroscopic problem
\eqref{eq:pb_macro} leads globally to the Lagrange multiplier
$\V{\widetilde{W}}^M$, which is finally used as a prescribed
displacement for the resolution of the substructure-independent
problems \eqref{eq:pb_micro}.

In order to carry out the resolutions of \eqref{eq:pb_micro} in
substructure variables, one uses the finite element method. Since
the constitutive law of the substructures is linear, the stiffness
operator of each substructure can be factorized once at the
beginning of the calculation and reused without modifications
throughout the analysis, which makes the method numerically
advantageous.\\

Algorithm~\ref{alg:LaTIn} summarizes the iterative procedure
described in this section.

\begin{algorithm2e}[ht]\caption{The two-scale domain decomposition solver}\label{alg:LaTIn}

Construction of each substructure's operators \;

Calculation of each substructure's macro homogenized behavior
$\mathbb{L}_E^M$ \;

Global assembly of the macroscopic operator\;

Initialization $s_0 \in \boldsymbol{\Gamma}$\;
\For{$n=0,\ldots,N$}{%
  Linear stage: calculation of $s_n \in \mathbf{A_d}$ \;
   $\quad \square$ Calculation of the macroscopic right-hand term $\widetilde{F}_E$ for each substructure \;
   $\quad \square$ Global assembly of the macroscopic right-hand term \;
   $\quad \square$ Resolution of the macro problem \;
   $\quad \square$ Resolution of the micro problems \;
  Local stage: calculation of $s_{n+\frac{1}{2}} \in \boldsymbol{\Gamma}$ \;
   $\quad \square$ Resolution of the local problems at the boundary interfaces $(\Gamma_{cl})_{E \in \mathbf{E}}$ \;
   $\quad \square$ Resolution of the local problems at the interfaces $(\Gamma_{EE'})_{(E,E') \in \mathbf{E}^2}$\;
  Calculation of a global error indicator
}
\end{algorithm2e}

\subsection{First example of a delamination analysis}
\label{sec:delam_examp}

\begin{figure}[htb]
       \centering
       \includegraphics[width=0.8 \linewidth]{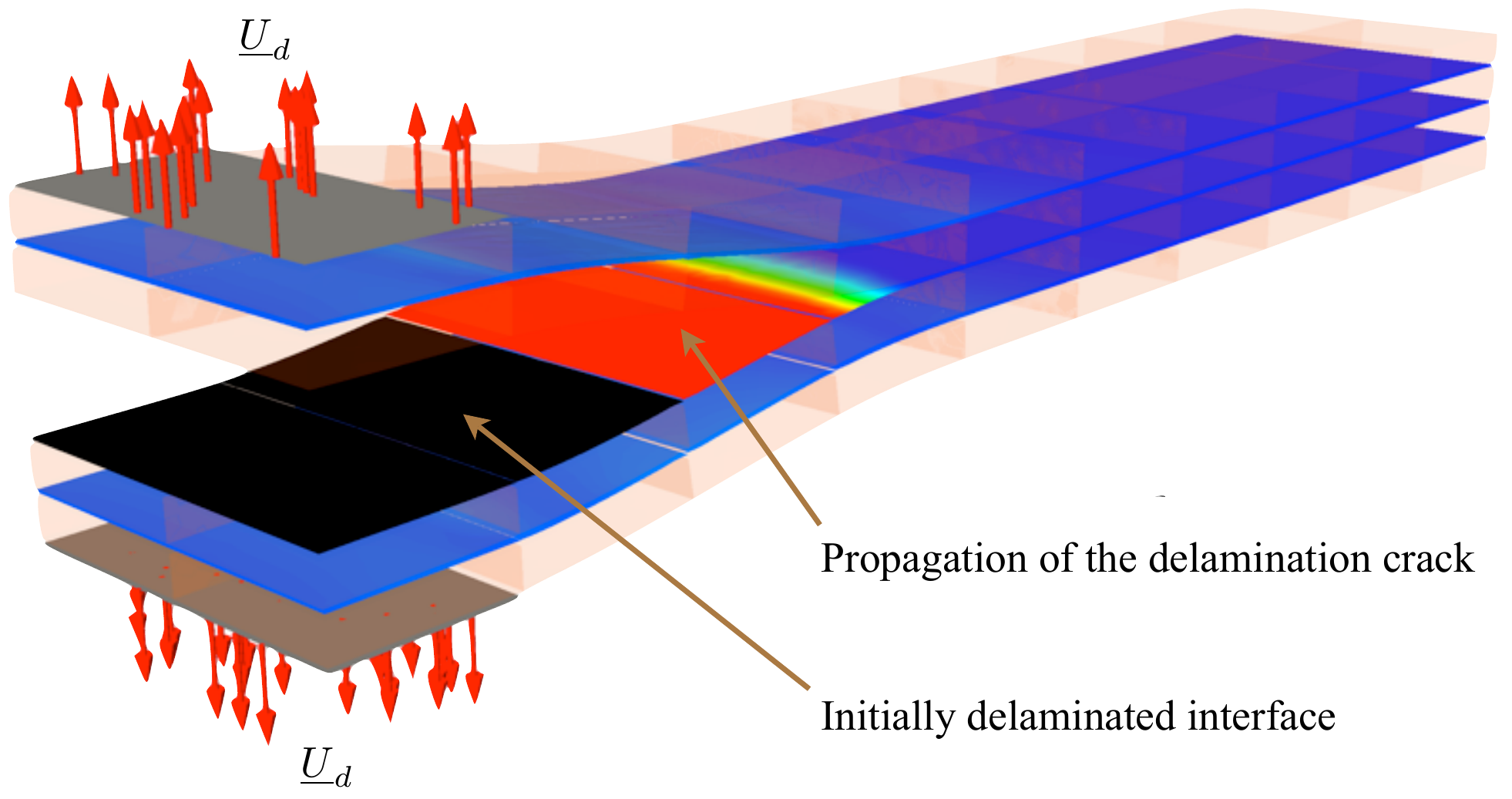}
       \caption{The four-ply DCB test example}
       \label{fig:DCB}
\end{figure}

A first example of quasi-static delamination analysis is shown in
Fig.~\ref{fig:DCB}. The structure is a $[0 \ 90]_s$ double
cantilever beam (DCB). The loading leading to Mode-I quasi-static
propagation of the crack is increased linearly through ten time
steps, the first two corresponding to the initiation of the
delamination and the remainder to the crack's propagation.
\ \\

The calculations were performed using a C++ implementation of the
mixed domain decomposition method capable of handling the
quasi-static analysis of 3D nonlinear problems. In this code, the
parallel computations use the MPI library to exchange data among
several processors. Each processor is assigned a set of connected
substructures and their interfaces; then it calculates the
associated operators and solves the local problems
(Fig.~\ref{fig:third_scale}). The allocation of the substructures
among the CPUs is handled by a METIS routine. The resolution of the
macroproblem does not take full advantage of the parallelism because
the substructures send their contributions to the macro problem to a
separate processor in which the matrix is assembled and factorized
and the substitutions are performed.
\ \\

\begin{figure}[htb]
       \centering
       \includegraphics[width=0.75 \linewidth]{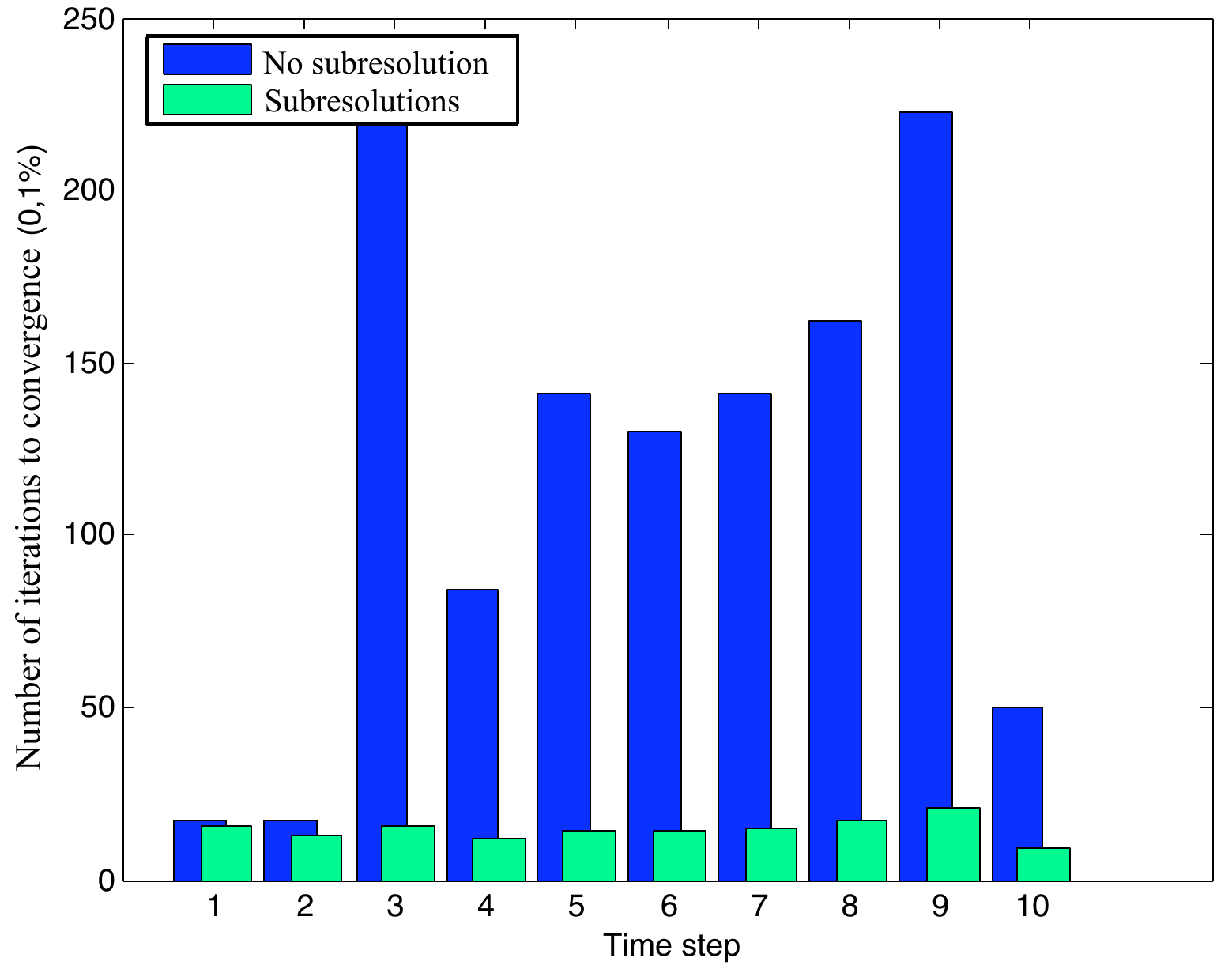}
       \caption{Subiterations near the crack's tip}
       \label{fig:sub_iterations}
\end{figure}

The direct use of the multiscale domain decomposition strategy to
simulate the DCB case led to a number of numerical difficulties:
\begin{itemize}
\item The convergence rate of the LATIN-based strategy is highly
dependent on the residual stiffnesses of the cohesive interfaces
as well as the values of the search direction parameters. The
iterative solver is even likely to stall when using low values
of the search direction parameters. In the next section, we will
briefly describe a practical tuning algorithm for the strategy
which guarantees convergence.
\item The method looses its numerical scalability when the crack's
tip propagates. This phenomenon appears clearly in
Fig.~\ref{fig:sub_iterations} under the label ``No
subresolution''. When the delamination process propagates (time
steps 3 to 10), the number of LATIN iterations required for
convergence becomes very large. A solution to this problem,
described in \cite{kerfriden09}, enabled us to recover the
scalability of the method for our test case (under the label
``Subresolutions''). This approach, which still needs to be
generalized, is based on a filtering of the long-range effects
of the crack's tip achieved through the resolution, at each
global LATIN iteration, of local nonlinear problems in a box
surrounding the front (where the main sources of nonlinearities
are located).
\item In this case, the ratio of the number of microscopic DOFs
to the number of macroscopic DOFs is relatively small (40). The
direct resolution of the macroscopic problem would become an
issue if one were addressing the simulation of a realistic
composite structure. A solution to this problem is discussed in
Section \ref{sec:third_scale}.
\end{itemize}

\section{Analysis of the parameters of the iterative algorithm}
\label{sec:ddr}
A necessary condition for the algorithm to converge is for the
search direction parameters $k^+$ and $k^-$ to be positive definite,
symmetrical operators. Previous studies have shown that there exists
an optimum set of these operators. However, the optimum values are
known to be difficult to interpret when the interface constitutive
laws are complex, and even in simplified cases (perfect interfaces)
are expensive to calculate. Therefore, our objective was to derive
an efficient scalar approximation of these search direction
operators for debonding analysis purposes.

As explained in \cite{kerfriden09}, the non-monotonic relation
between the interface stresses and the displacement gap due to
damage imposes restrictions on the choice of Parameter $k^+$.
Concerning $k^-$, optimum values for a given damage map can be found
after a micro/macro decomposition of this operator. Since the status
of the interfaces changes with the evolution of the delamination
(from elastic to damaged, then from damaged to ruined), the search
direction parameters need to be updated often in order to remain
optimum. Therefore, parameters whose efficiency range when they are
not optimum is broad enough to require less frequent updating are
preferred. An effective practical choice is the following:
\begin{itemize}
\item Parameter $k^+$: in order to avoid stalling or divergence
of the algorithm, this parameter is set to a very high value
(\textit{i.e.} the search direction $E^+$ is quasi-infinitely
stiff).
\item Parameter $k^-$:
\begin{itemize}
\item perfect interfaces: $k^-$ is set to the classically recommended
value $E/L$  \cite{ladeveze03b}, where $E$ is the Young's
modulus of the adjacent substructure and $L$ a
characteristic length of the interface.
\item interfaces with prescribed forces (respectively displacements):
$k^-$ is set to a very small (respectively large) value in
order to enforce the boundary condition through penalization
in the adjacent substructure.
\item cohesive interfaces: $k^-$ is set to the stiffness of the
undamaged interface.
\item delaminated interfaces: $k^-$ is set to zero in the shear direction.
In the normal direction, $k^-$ is set to zero in traction
and to the initial stiffness in compression. Therefore, the
status of the interface must be checked regularly (e.g.
every ten iterations).
\end{itemize}
\end{itemize}
The use of an infinitely stiff search direction $E^+$ and of the
initial cohesive interface stiffness as the search direction
parameter $E^-$ brings us back to a well-known situation. The
algorithm can be viewed as a secant Newton algorithm in which the
solutions of the prediction steps are in equilibrium only in the
macroscopic space, the equilibrium of the microscopic quantities
being achieved at convergence.

\section{The three-scale domain decomposition strategy}
\label{sec:third_scale}

\begin{figure}[htb]\centering
\subfigure[Stresses and displacements in the substructures]{\includegraphics[width=0.9 \linewidth]{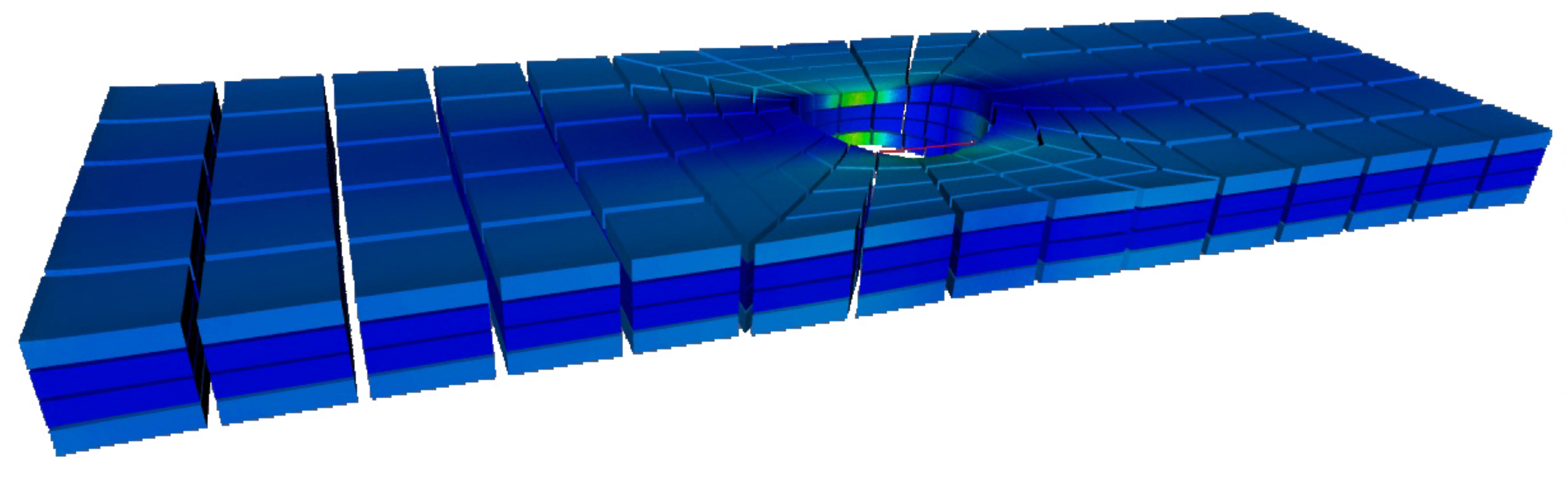}}
\subfigure[Damage over the upper interface]{\includegraphics[width=0.9 \linewidth]{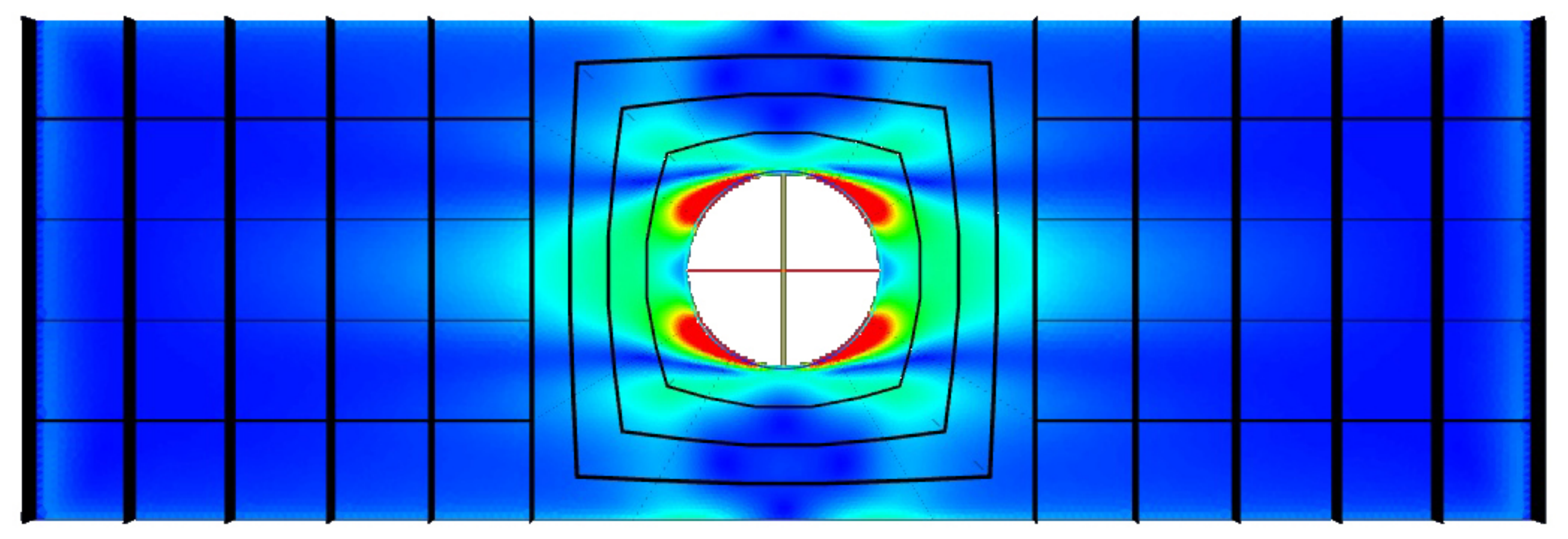}}
\caption{The four-ply perforated plate problem (3.4 MDOFs)}
\label{fig:holed}
\end{figure}

The decomposition into substructures described in
Section~\ref{sec:reference_problem} leads to a very large macro
problem and an unnecessarily refined macroscopic solution. In order
to solve large problems such as that represented in
Fig.~\ref{fig:holed}, one must place the emphasis on the parallel
resolution of the macroproblem and on the selection and transmission
of the large-wavelength part of the macroscopic solution.

These two features can be introduced into the method by using any
Schur-complement-based domain decomposition technique
\cite{gosselet06}. We chose to solve the macroproblem using the BDD
method \cite{mandel93,letallec94}.

\subsection{Resolution of the macroproblem through the balancing domain decomposition method}

\subsubsection{Partitioning of the macroproblem}

\begin{figure}[htb]\centering
       \includegraphics[width=0.8 \linewidth]{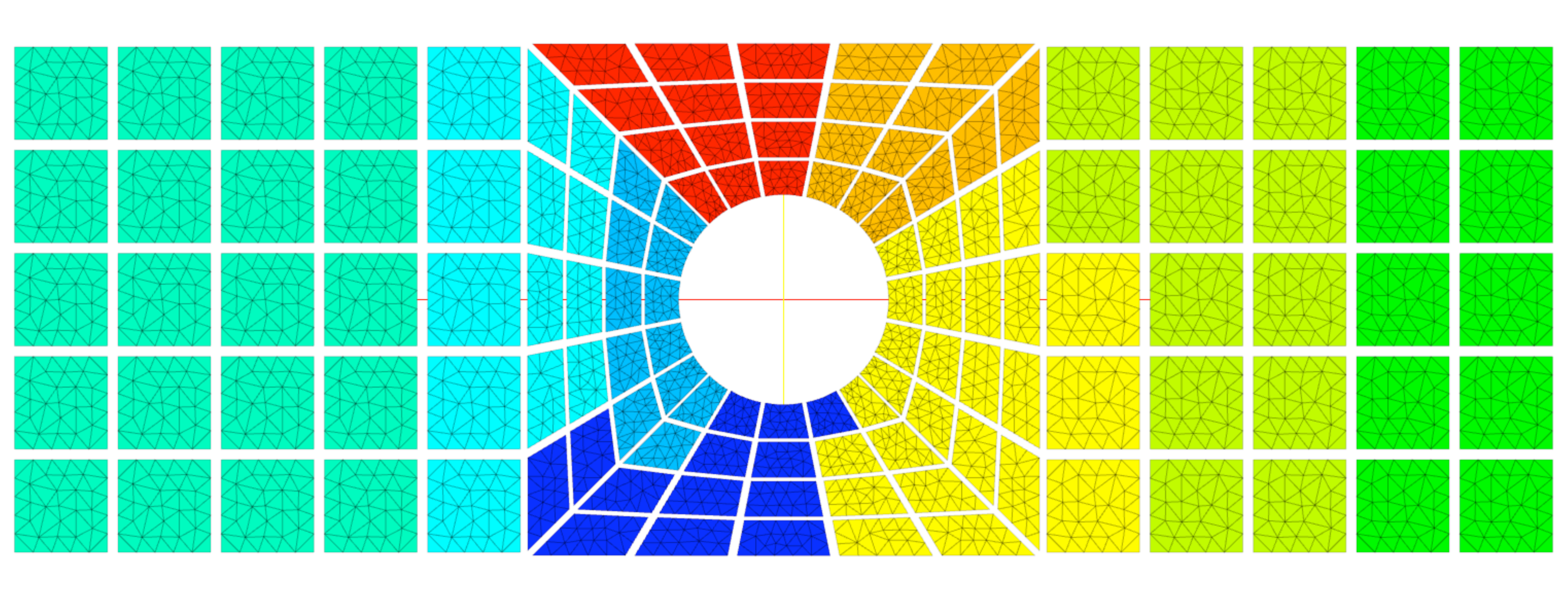}
       \caption{Three-level substructuring: assignment of substructures to processors}
       \label{fig:third_scale}
\end{figure}

The substructures of the initial partitioned problem are grouped
into super-substructures ($\bar{E}$) separated by super-interfaces
$\Gamma_{\bar{E}\bar{E'}}$ (Fig.~\ref{fig:third_scale}). The
algebraic problem to be solved within each of these
super-substructures (dropping Superscript $M$) is:

\begin{equation}
\label{eq:decomp}
\left\{ \begin{array}{c}
\left( \begin{array}{cc}
\displaystyle \mathbf{L}_{ii}^{(\bar{E})} & \displaystyle \mathbf{L}_{ib}^{(\bar{E})} \\
\displaystyle \mathbf{L}_{bi}^{(\bar{E})} & \displaystyle \mathbf{L}_{bb}^{(\bar{E})}
\end{array} \right)
\left( \begin{array}{c}
\displaystyle \widetilde{W}_i^{(\bar{E})} \\
\displaystyle \widetilde{W}_b^{(\bar{E})}
\end{array} \right)
=
\left( \begin{array}{c}
\displaystyle F_i^{(\bar{E})} \\
\displaystyle F_b^{(\bar{E})}+ \widetilde{\lambda}_b^{(\bar{E})}
\end{array} \right)
\\
\widetilde{W}_b^{(\bar{E})} = {\mathbf{A}^{(\bar{E})}}^T\V{\widetilde{W}}_b
\\
\displaystyle \sum_{\bar{E}}  \mathbf{A}^{(\bar{E})}  \widetilde{\lambda}_b^{(\bar{E})} = 0
\end{array} \right.
\end{equation}
where Subscripts $b$ and $i$ refer respectively to the
super-interface quantities and to the internal quantities of the
super-substructures. $\mathbf{A}^{(\bar{E})}$ is a Boolean operator
which localizes data in such a way that the second equation of
System \eqref{eq:decomp} expresses the continuity of the kinematic
unknowns (deduced from a single unknown $\V{\widetilde{W}}_b$),
while the third equation expresses the equilibrium of the nodal
reactions at the super-interfaces.

First, the local equilibrium is condensed onto the super-interfaces
by introducing the Schur complement $\mathbf{S}^{(\bar{E})}$ and the
condensed force $\V{F}_c^{(\bar{E})}$. The assembled condensed
problem becomes:
\begin{equation}
\mathbf{S} \  \V{\widetilde{W}}_b = \V{F}_c
\end{equation}
\begin{equation*}
\textrm{where} \quad
\left\{ \begin{array}{l|l}
\displaystyle \mathbf{S} = \sum_{\bar{E}}  \mathbf{A}^{(\bar{E})} \mathbf{S}^{(\bar{E})} {\mathbf{A}^{(\bar{E})}}^T \V{\widetilde{W}}_b\ & \displaystyle \mathbf{S}^{(\bar{E})} = \mathbf{L}_{bb}^{(\bar{E})} - \mathbf{L}_{bi}^{(\bar{E})} \  \mathbf{L}_{ii}^{{(\bar{E})}^{-1}} \ \mathbf{L}_{ib}^{(\bar{E})}\\\
\displaystyle  F_c = \sum_{\bar{E}}  \mathbf{A}^{(\bar{E})} \V{F}_c^{(\bar{E})} & \displaystyle F_c^{(\bar{E})} = F_b^{(\bar{E})} - \mathbf{L}_{bi}^{(\bar{E})} \  \mathbf{L}_{ii}^{{(\bar{E})}^{-1}} F_i^{(\bar{E})}
\end{array} \right.
\end{equation*}


This substructuring technique can be used exactly as in
\cite{ramm.2008.1} in order to bind a domain which is prone to
localization and damage to an undamaged region. Nevertheless, for
large interface problems such as those encountered in our case, the
condensed problem is much too large to be solved directly, and
iterative solvers must be used.

\subsubsection{Resolution of the super-interface problem}

The condensed macroproblem is solved iteratively using a conjugate
gradient algorithm. Classically, this resolution involves only
matrix-vector products and dot products, which are compatible with
parallel computation.
The recommended Neumann-Neumann preconditioner
$\mathbf{\widetilde{S}}^{-1} $ involves the use of the
pseudo-inverses ${\mathbf{S}^{(\bar{E})}}^{+}$ of the Schur
complements of the super-substructures:
\begin{equation}
\mathbf{\widetilde{S}}^{-1} =
\sum_{\bar{E}} \mathbf{A}^{(\bar{E})} {\mathbf{S}^{(\bar{E})}}^{+} {\mathbf{A}^{(\bar{E})}}^T
\end{equation}
The use of this preconditioner means that the inverse of the global
super-macro operator is approximated by the assembly of the inverses
of the local Schur complements. Let us note that the description
chosen for the interface macrofields precludes the existence of
degrees of freedom belonging to more than two substructures;
consequently, no scaling is required in the preconditioner (at least
as long as the interfaces are not excessively heterogeneous). The
use of pseudo-inverses is associated with an optimality condition
which ensures that rigid body motions are not solicited
(self-equilibrium of the floating super-substructures). This
condition is verified thanks to a projector which makes the residual
orthogonal to the kernels of the super-substructures (and possibly
to other given subspaces) at each iteration of the conjugate
gradient.

\subsection{Results}

\begin{figure}[ht]
       \centering
       \includegraphics[width=0.75 \linewidth]{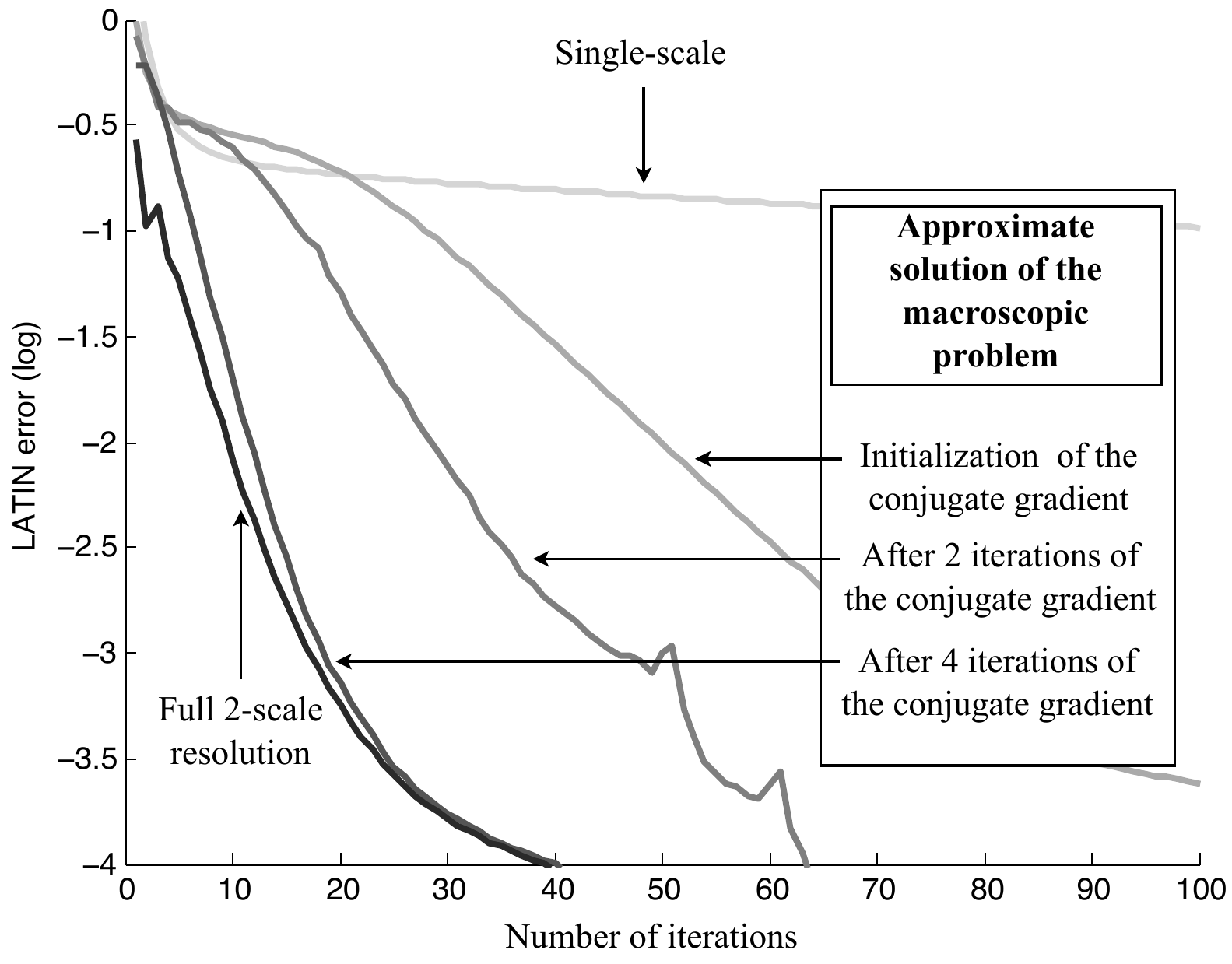}
       \caption{The LATIN convergence curves (error criterion \textit{vs.}
       the number of iterations) for several numbers of macroiterations}
       \label{fig:cg_conv}
\end{figure}

Fig.~\ref{fig:cg_conv} shows the convergence rate of the LATIN
algorithm when the conjugate gradient algorithm for the condensed
macroproblem is stopped after a fixed number of iterations. The test
case was the perforated plate under traction represented in
Fig.~\ref{fig:holed} with the decomposition into super-substructures
of Fig.~\ref{fig:third_scale}. It is clear that a rough
approximation of the Lagrange multiplier, obtained after very few
iterations of the conjugate gradient, is sufficient to reach the
convergence rate of the multiscale LATIN algorithm.
Typically,
the algorithm is stopped when the residual error (normalized by the
initial error) falls below $10^{-1}$. Thus, the third-level
enforcement of the admissibility of the macroforces (through the
projection) appears to be sufficient for the determination of the
large-wavelength part of the solution to be transmitted through the
structure at each iteration of the resolution.

\section{Efficiency of the strategy: study of a complex test case}
\label{sec:complex_test}
\begin{figure}[p]
       \centering
       \includegraphics[width=0.99 \linewidth]{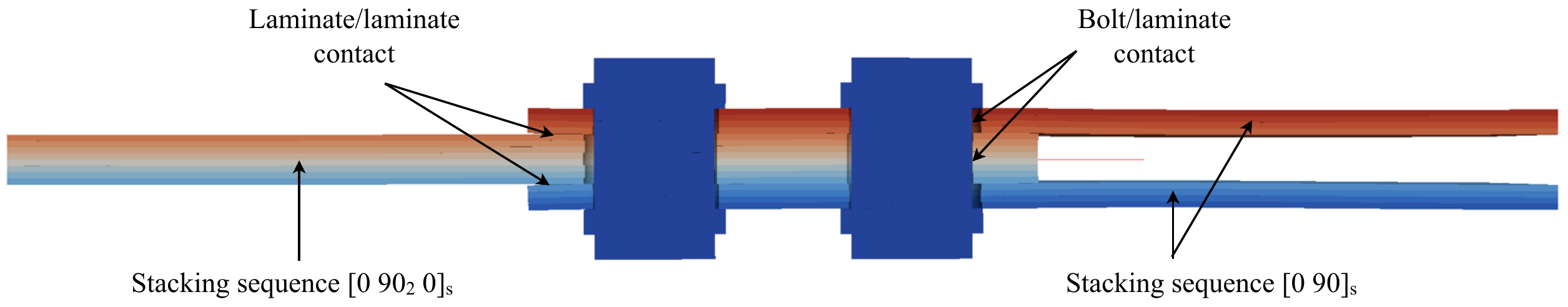}
       \caption{Composite bolted joint made of 16 $0.125$ mm-thick plies.
       The in-plane dimensions of the structure are $30 \times 5$ mm.
       Prescribed displacements are applied along the left-hand side of the
       $[0 \ 90_2 \ 0]_s$ composite plate and along the right-hand sides of
       the  $[0 \ 90]_s$ composite plates.}
       \label{fig:liaison_def}
\end{figure}

In this section, we illustrate the efficiency of the three-scale
domain decomposition strategy through the simulation of the
evolution of debonding in the bolted composite joint shown in
Fig.~\ref{fig:liaison_def}. Each composite plate interacts with the
adjacent plates and with the two steel bolts through contact
interfaces. The structure is subjected to prescribed displacements
along the edges of the plates.

\begin{figure}[p]
       \centering
       \includegraphics[width=0.99 \linewidth]{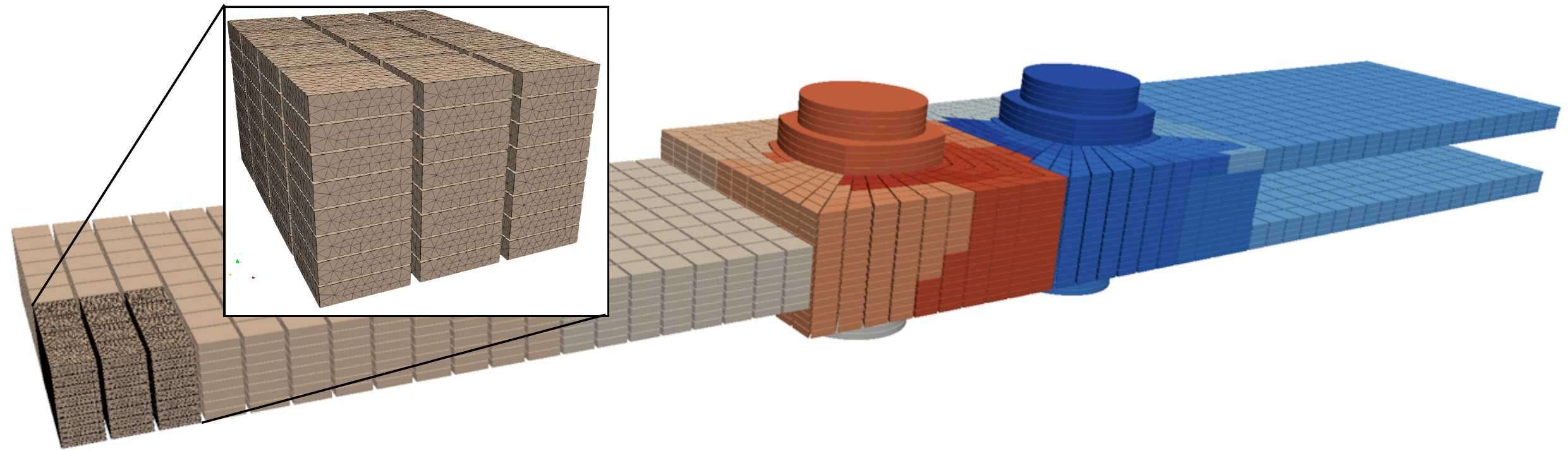}
       \caption{Discretization of the composite bolted joint ($12\, 10^6$
       DOFs), decomposition into substructures ($10,600$ substructures)
       and assignment to processors (29 CPUs)}
       \label{fig:liaison_sst}
\end{figure}

\begin{figure}[p]
       \centering
       \includegraphics[width=0.99 \linewidth]{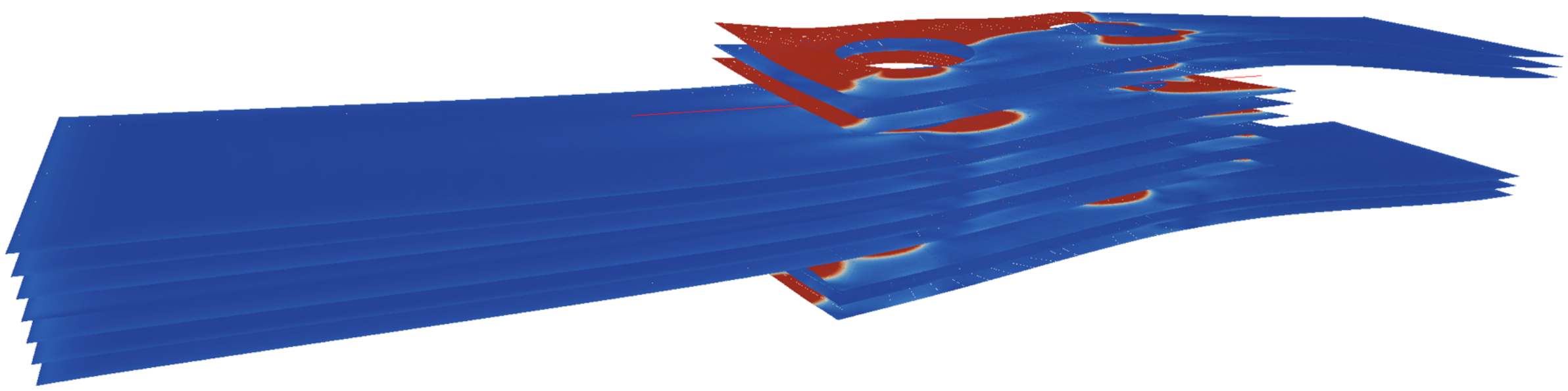}
       \caption{Damage map of the cohesive interfaces of the composite
       bolted joint after the 70$^{\textrm{th}}$ time step of the quasi-static
       incremental analysis procedure}
       \label{fig:liaison_damage}
\end{figure}

The discretization and the decomposition into substructures for this
test case are illustrated in Fig.~\ref{fig:liaison_sst}. The total
number of DOFs involved was $12\, 10^6$, distributed among $10,600$
substructures. The number of macroscopic DOFs was $3 \, 10^5$, which
would have made the direct resolution on a standard computer very
inefficient. 29 processors each with 4 Gigabytes memory were used
for this calculation, which led to a super-coarse grid problem of
dimension 150 (6 unknowns per floating super-substructure).

Figure~\ref{fig:liaison_damage} shows the damage map in the
composite bolted joint after 70 time steps. The nonlinear
calculation corresponding to each time step was carried out until
the LATIN error criterion got below $10^{-2}$, which occurred after
an average of $80$ LATIN iterations per time step. The average CPU
time required for the calculation of each time step was 30 minutes,
which is reasonable considering the small number of processors used.

However, the number of global LATIN iterations was quite high
compared to what could have been obtained using the relocalization
strategy of \cite{kerfriden09} in the vicinity of the crack's front
(see the results of Fig.~\ref{fig:sub_iterations}).
Figure~\ref{fig:liaison_damage} is a clear illustration of the
difficulties which may arise in the use of this dedicated technique
in a general case in which multiple crack front propagations may be
involved. The front has a complex shape, which raises the difficult
issue of the choice of the number of relocalization zones and their
sizes. In addition, this test case is very unstable. These
instabilities were handled globally using an arc-length algorithm
along with the three-scale resolution strategy, but local
instabilities might also appear within the region extracted for the
relocalization calculations, a situation which has not yet been
addressed at this stage in our development. Therefore, in the
future, it might be necessary to generalize the relocalization
strategy in order to improve the efficiency of the enhanced
multiscale domain decomposition technique for complex laminated
structures.

\section{Conclusion}

The accurate prediction of delamination in extended process zones of
laminated composite structures requires refined models of the
material's behavior, leading to the resolution of huge systems of
equations. In order to solve such problems accurately, we used a
two-scale domain decomposition strategy based on an iterative
resolution algorithm. This method is particularly appropriate for
laminated mesomodels, in which 3D and 2D entities are introduced
separately.

This strategy was improved in order to enable it to handle very
large delamination problems. A systematic analysis of the features
of the method on the different scales was performed. First, we
showed that in the high-gradient zones the classical scale
separation was insufficient to ensure numerical scalability.
Therefore, we developed a subresolution procedure which preserves
the numerical scalability of the crack propagation analysis, but
still needs to be automated for complex structures and regulated
against local instabilities. We also proved that a third scale is
required. Then, the problem on the intermediate scale was solved
using a parallel iterative algorithm which enabled the rapid
transmission of the very-large-wavelength part of the solution.
Global instabilities were handled through a classical arc-length
algorithm with local control (e.g. based on the maximum damage
increment) and adjustment of the ``time'' steps during the
calculation of the evolution of damage.

In future developments, 3D analysis in the process zone will be used
in conjunction with plate analysis, which would be sufficient to
describe the solution in the low-gradient zones. We will also,
within the MAAXIMUS project, investigate the interaction between
delamination and buckling for the simulation of components of
aeronautical structures.

\bibliographystyle{plain}
\bibliography{template}


\end{document}